\documentclass[aip,
amsmath,amssymb,
reprint
]{revtex4-1}

\usepackage{graphicx}% Include figure files
\usepackage{dcolumn}% Align table columns on decimal point
\usepackage{bm}% bold math
\usepackage{soul}

\newcommand{\Temp}{k_{\mathrm{B}}T}
\usepackage{afterpage}

\usepackage{color}
\begin{document}

\preprint{AIP/123-QED}

\title{Kinetic Monte Carlo Simulations for Birefringence Relaxation of Photo-Switchable Molecules on a Surface }

%\title{Kinetic Monte Carlo Simulation for the Rotational Dynamics of Photo-Switchable Molecules on a Surface.}

\author{Raffaele Tavarone}
\email{raf.tav@mail.tu-berlin.de }
\affiliation{Institut f\"ur Theoretische Physik, Technische Universit\"at Berlin,
Hardenbergstrasse 36, D-10623 Berlin, Germany}

\author{Patrick Charbonneau}
%\email{patrick.charbonneau@duke.edu}
\affiliation{Departments of Chemistry and Physics, Duke University, Durham, North Carolina 27708, USA}

\author{Holger Stark}
\email{holger.stark@tu-berlin.de}
\affiliation{Institut f\"ur Theoretische Physik, Technische Universit\"at Berlin,
Hardenbergstrasse 36, D-10623 Berlin, Germany}

%


\begin{abstract}
Recent experiments have demonstrated that in a dense monolayer of photo-switchable dye Methyl-Red 
molecules the 
relaxation of 
an initial
%the}
birefringence
follows a power-law decay, typical for glass-like dynamics.
%a glasslike,} decays as a} 
%power law \pc{\st{dynamics}}. 
The slow relaxation can efficiently be controlled and 
%fastened
accelerated by illuminating
the monolayer with circularly polarized light, which induces \textit{trans}-\textit{cis} isomerization cycles.
To elucidate the microscopic mechanism, we 
develop a two-dimensional molecular model in which the \textit{trans}
and \textit{cis} isomers are 
represented by straight and bent needles, respectively.
As in the experimental system, the needles are allowed to rotate 
and to form overlaps
but 
they
cannot translate.
The 
out-of-equilibrium rotational dynamics 
of the needles
is 
generated
using kinetic Monte Carlo simulations.
We demonstrate that, in a regime of high density and low temperature, the power-law relaxation can be traced 
to the 
formation
of spatio-temporal correlations in the rotational dynamics, \textit{i.e.}, 
dynamic
heterogeneity. We also show that the nearly 
isotropic \textit{cis} isomers can  
prevent dynamic
heterogeneity
from forming in the monolayer and 
that the relaxation then  
becomes exponential.  
%Valid PACS numbers may be entered using the \verb+\pacs{#1}+ command.
\end{abstract}

%\pacs{42.70.Gi}% PACS, the Physics and Astronomy
\maketitle

\section{Introduction}

The possibility to control organic and inorganic materials at the molecular nanoscale level is 
crucial for a large variety of technological application and for a deeper understanding of matter \cite{fang2013athermal,grier2003revolution,hanggi2009artificial,yan2009nanowire,craighead2006future,ritort2006single,caruthers2007nanotechnological,cao2002fabrication}. Among 
possible 
tools for
molecular control, light is one of the most promising. 
Some of the
appealing
applications
include: 
illuminating the metallic tip of a scanning force microscopy 
%\pc{\st{allows}} 
to precisely control the position of single molecules \cite{grier2003revolution}, 
using nanowires 
to build miniaturized photonic devices \cite{yan2009nanowire}, 
and inscribing nano-sized geometrical patterns 
on a surface by photolithography \cite{shao2005surface}.

Photochromic molecular switches, molecules that
undergo configurational changes
between two (or even more) isomeric states
when irradiated by light \cite{delaire2000linear},
offer yet another appealing 
way to control material properties with light. 
For example, they are
%for instance} 
used to fabricate
functional surfaces with tunable chirality, wettability, conductivity etc. \cite{katsonis2007synthetic,browne2009light}.
Among other things \cite{kreuzer2000light},
photoswitching molecules illuminated by light can reorient a nematic liquid crystal or
directly control both the 
formation
and 
relaxation of orientational order in 
a
monolayer \cite{kosa2012light,karageorgiev2005anisotropic,fang2013athermal,fang2011effect,fang2010photo}.  
Thus, many 
%Many} 
material properties such as mass transport, mobility, and viscosity can be 
efficiently tuned
%with 
%\pc{\st{high efficiency and completely} 
while producing
%negligible 
hardly any
%\pc{\st{production of}}
heat.
Although
these phenomena  have attracted much interest for many years, 
%\pc{\st{but still}} 
details of the microscopic dynamics
still need to be clarified.

In theory the rotational dynamics of light-switchable molecules is mainly described by Fokker-Planck equations 
for the molecular orientational distribution functions of each isomeric configuration supplemented by source terms
\cite{janossy1994molecular,marrucci1997photoinduced,janossy1998optical,pedersen1997mean,chigrinov2004diffusion,sekkat1995reorientation,statman2003study,kiselev2002kinetics,chen1996model,marrucci1997role}.
Molecular interactions are treated within mean-field approximation, where microscopic details such as the different 
%\pc{isomer} 
shapes 
of the isomers
are neglected.

In this article we consider a self-assembled monolayer of  light-switchable molecules tethered to a surface. Instead of the method 
mentioned above, we perform kinetic Monte Carlo simulations for a molecular model, where we approximate the two isomeric states,
called \emph{trans} and \emph{cis}, by a straight and a bent needle, respectively. The simplicity of the model allows us to  study 
the long-time collective dynamics of a statistical ensemble consisting of 10,000 molecules \cite{battaile2008kinetic}, 
much more than atomistic molecular dynamics simulations
can handle.
In a previous work we have investigated the phase ordering of bent needles with varying shape \cite{tavarone2015phase}.

Our work is motivated by a recent experimental study of Fang \emph{et al.} \cite{fang2013athermal}
on the glasslike orientational dynamics of a self-assembled monolayer of photo-switching molecules.
After aligning the molecules with 
light, the authors observed the decay of orientational order (or birefringence)
under either thermal erasure or erasure with circularly polarized (CP) light. 
In both cases they find that the relaxation of birefringence follows a power law,
which is typical for glasslike dynamics.

Within our relatively simple model we can reproduce this feature in a system 
%\pc{\st{only consisting of} 
containing straight needles alone 
(\emph{trans} molecules), if the density is sufficiently high and temperature is low. We demonstrate that the needles, when 
randomizing their orientations, develop dynamic heterogeneities in space and time \cite{berthier2011phys,berthier2011theoretical}, 
which ultimately cause the power-law decay. The presence of \emph{cis} molecules, which have a rather isotropic shape, can 
prevent the formation of such spatio-temporal variations in the local structure and the birefringence relaxation then
becomes exponential. In the following, we clarify under which conditions our model nevertheless reproduces the experimental 
observation of a power-law decay by tuning isomerization probabilities.

The plan of this paper is as follows.
In Sec.\ \ref{sec: experiment} we thoroughly review the experimental motivation for our work and explain the model and simulation
method in Sec.\ \ref{sec: Model and Method}. The system of pure \emph{trans} molecules is studied in detail in 
Sec.\ \ref{sec: needles relaxation} and in
Sec.\ \ref{sec: relaxation isomerization} we discuss
the birefringence relaxation in a system with light-switchable molecules. We close with a summary of the results
and a conclusion in Sec.\ \ref{sec: conclusion}.

% % % % % % % % % % % % % % % % % % % % % % % % % % % % % % % % % %
% % % % % % % % % % % % % % % % % % % % % % % % % % % % % % % % % %
% %
% % Section: Experimental motivation
% %
% % % % % % % % % % % % % % % % % % % % % % % % % % % % % % % % % %
% % % % % % % % % % % % % % % % % % % % % % % % % % % % % % % % % %

\section{\label{sec: experiment} Experimental Motivation}

The work presented in this paper is strongly inspired by recent
experiments of Fang \emph{et al.} \cite{fang2013athermal} that have
attracted considerable attention. In this section we first shortly summarize their results. 

A self-assembled monolayer (SAM)
with glasslike dynamics
is realized by covering at high in-plane density a glass surface with dye Methyl-Red (dMR) molecules  
and tethering them with covalent bonds at random positions.
The molecules are then free to rotate but their translational freedom is constrained to $\approx 1$nm displacements. 
In the ground state dMR molecules  
assume a rod-like \textit{trans}
configuration with anisotropic optical properties. The light-induced
\textit{cis} configuration has a bent-core shape
and is nearly isotropic. 
Because of their photo-switchable azo-core, isomerization between the two configurations
can efficiently be induced by illumination with light
at a 514 nm wavelength. 
The absorption spectra of the \textit{cis} and \textit{trans} isomers 
partially overlap and illumination with light of  this wavelength 
induces both \textit{cis}-to-\textit{trans} and \textit{trans}-to-\textit{cis} transitions.

In Ref.~[\onlinecite{fang2013athermal}] the initial orientation of the molecules in
the SAM was random. Under illumination with linearly polarized light (writing process), 
molecules preferentially aligned along the light polarization vector more likely switch toward the
\textit{cis} configuration, while the remaining \textit{trans} molecules form nematic order perpendicular to the light polarization
(hole-burning) \cite{kiselev2009kinetics,yi2008high,chigrinov2008photoalignment}.
The monolayer thus exhibits some
birefringence $Q(t)$.
Since 
light illumination also induces 
the reverse,  \textit{cis}-to-\textit{trans}
transition,  
in steady state
a mixture of both isomer exists  at a relative concentration
that mostly depend on the light intensity and the monolayer density.
%\pc{\st{of the monolayer and other factors.}}

Reference [\onlinecite{fang2013athermal}] investigated the relaxation of the birefringence under two different illumination conditions 
following the writing process:
\newline
\textbf{1)} The SAM is left in the dark, resulting in the relaxation being driven by thermal fluctuations (thermal erasure). 
At room temperature all molecules in the \textit{cis} configuration 
relax back to the \textit{trans} form after a characteristic time \cite{yi2011dynamics}. They assume random orientations
and the nematic order is lost.
\newline
\textbf{2)} The SAM, immediately after the end of the writing process, is instead illuminated with circularly polarized light 
at 514 nm, which makes the molecules cycle between the \textit{trans} and \textit{cis} configurations 
and speeds
up the relaxation dynamics of the birefringence (CP erasure).

\begin{figure}
\begin{center}
\includegraphics{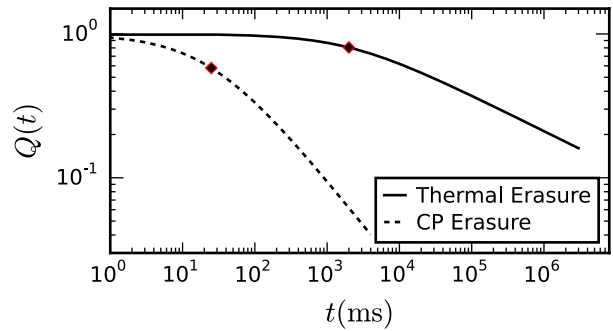}
\caption{Schematic representation of the relaxation of the birefringence, $Q(t)$, 
obtained in the experiment discussed in Ref.~\onlinecite{fang2013athermal}.
The birefringence inscribed in the monolayer by illuminating it with linearly polarized light can be 
erased either by thermal fluctuations at room temperature (thermal erasure) or by illumination with 
circularly polarized light (CP erasure). In both cases the relaxation is well described by 
the long-time power law decay 
%\pc{form} 
of Eq.~(\ref{eq: PSSG relaxation}). 
Parameters $\tau_t$ are shown as a diamond marker. 
Other parameters are $\eta=0.25$ for the thermal erasure curve and $\eta=0.63$ for the CP erasure curve.  }
\label{fig: experiment scheme}
\end{center}
\end{figure}

A schematic of the typical experimental results is given in 
Fig. \ref{fig: experiment scheme}, where the temporal evolution of the birefringence $Q(t)$ is shown for both thermal and CP erasure processes.
Fang \emph{et al.}\cite{fang2013athermal} 
%also 
showed that the relaxation of the birefringence is non-exponential, and is instead 
accurately described by a functional form of the type
\begin{equation}
Q(t)=[1+(t/\tau_t)]^{-\eta} \, .
\label{eq: PSSG relaxation}
\end{equation} 
An
asymptotic power-law relaxation  
starts
at $t\approx \tau_t$ and 
proceeds
as $Q(t)\approx (t/\tau_t)^{-\eta}$ for $t> \tau_t$.

Non-exponential relaxations are characteristic of 
a glassy state,  
i.e., of systems sampling a rugged free energy landscape\cite{shlesinger1988fractal,metzler2002stretched,vainstein2006non,bohmer1993nonexponential}.
The underlying dynamics does not possess a single characteristic 
time scale, but rather a distribution of them.
It has been suggested \cite{fang2013athermal,fang2011effect} that the 
distribution of energetic barriers that leads to the power-law relaxation in the SAM originates from 
i) the high packing density and 
ii) the molecular rotations proceeding in discrete jumps as molecules pass each other by stretching or squeezing 
the covalent bonds.  

In the following we describe a molecular model, which incorporates the essential features of the experiment 
outlined in this section and consider its dynamical behavior under conditions that mimic those studied experimentally
\cite{fang2013athermal}.

% % % % % % % % % % % % % % % % % % % % % % % % % % % % % % % % % %
% % % % % % % % % % % % % % % % % % % % % % % % % % % % % % % % % %
% %
% % Section: Model
% %
% % % % % % % % % % % % % % % % % % % % % % % % % % % % % % % % % %
% % % % % % % % % % % % % % % % % % % % % % % % % % % % % % % % % %

\section{\label{sec: Model and Method}Model and Simulation Methods}

In this section we detail the molecular model and
the kinetic Monte Carlo simulations that we implemented to investigate the relaxation of the birefringence in the 
experimental system.

\subsection{\label{sec: molecular model} Molecular model}
\begin{figure}
\begin{center}
\includegraphics{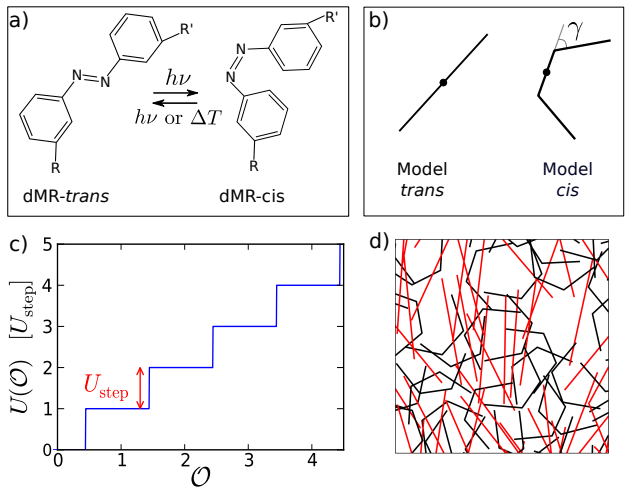}
\caption{(colors) (a) The dye methyl red (dMR) molecule
used in the experiments of
Ref. [\onlinecite{fang2013athermal}]. 
The azo-core of the molecule is responsible for the 
light-switching process 
%but 
and
other chemical groups are attached 
in the positions denoted by $\mathrm{R}$ and $\mathrm{R^{\prime}}$.
(b) The two dMR isomers are modeled by 
an infinitely thin needle for the \textit{trans} 
configuration and a bent version for the \textit{cis}
molecule.
(c) The step-potential $U(\mathcal{O})$ gives the energy of a molecular configuration as a function of the number 
of overlaps $\mathcal{O}$ that a molecule forms with its neighbors. 
All the steps have a fixed height 
$U_{\mathrm{step}}$. 
(d) A small fraction of a typical initial configuration 
of the kinetic Monte Carlo simulation showing \textit{cis} (black) and \textit{trans} (red) 
model molecules attached to a two-dimensional surface.}
\label{fig: molecular model}
\end{center}
\end{figure}

As explained in Sec.~\ref{sec: experiment}, when the monolayer of 
tethered
molecules studied in Ref.~[\onlinecite{fang2013athermal}] is 
illuminated by 
light,
molecules cycle between an anisotropic, rod-like \textit{trans} configuration and a nearly isotropic, bent-like \textit{cis} 
configuration [see Fig.\ \ref{fig: molecular model}(a)]. 
We model the \textit{trans} isomer as an infinitely thin, straight needle of unit length $L=1$ and 
the \textit{cis} isomer as a bent, infinitely thin needle also of total length $L$ 
[see Fig. \ref{fig: molecular model}(b)]. 
In the \textit{cis} configuration we fix the 
angle between the central and the tail segments to $\gamma=\pi/3$, 
while  each tail segment  
has a length of $0.35 L$.
%\rt{A previous work \cite{tavarone2015phase} on the phase ordering of these model-molecules in two-dimensions demonstrated 
%that, due to the presence of the external segments, bent needles order in a quasi-nematic phase at much higher packing density 
%than straight needles.}
The total number of \textit{trans} ($N_{\mathit{t}}$) and \textit{cis} ($N_{\mathit{c}}$)
molecules
is fixed, i.e., $N_{\mathrm{t}}+N_{\mathrm{c}}=N$. 
Since
the transition moment of the dMR molecules, to which the light polarization couples, 
is nearly parallel to the monolayer surface, we 
consider
the system to be purely two-dimensional. To mimic the 
effect of the covalent tethering, molecules are allowed to rotate 
within the plane but cannot translate.

The isomerization of an azobenzene is a very complicated process when considered in full atomistic detail.
The transition between the two isomeric states occurs via a number of intermediate excited states while several conformational 
degrees of freedom of the molecule change \cite{brzozowski2001azobenzenes,cusati2011photodynamics}.
The situation is even more complicated when 
the isomerization  
occurs in a  
crowded environment 
because
neighboring molecules then 
most likely interfere 
%\pc{\st{with it}}
\cite{tiberio2010does}.

In our model the isomerization process is drastically simplified: it consists of a simple switch from the straight to the 
bent needle and vice-versa. 
Also, we assume that the conformational change happens instantaneously, a reasonable assumption given the
relative time scales involved. 
The relaxation of the birefringence happens at least on the second scale (see Fig.~\ref{fig: experiment scheme}), 
while the isomerization occurs on the picosecond scale even in relatively dense organic solvents \cite{tiberio2010does}. 

In order to incorporate  
isomerization  
as described in Sec.~\ref{sec: experiment} in the kinetic Monte Carlo simulation, we define the following set of rules:
\newline
\textbf{1) thermal erasure}: only the $\mathit{cis} \rightarrow \mathit{trans}$ spontaneous transition is allowed and  
isomerization occurs with probability $P_{\mathrm{th}}(\mathit{c}\rightarrow \mathit{t})$. 
\newline
\textbf{2) CP erasure}:  under illumination 
with circularly polarized light the isomerization rate is proportional to the light intensity,
the quantum yield of the transition process, and the absorption cross sections of the isomers \cite{statman2003study,sekkat1995reorientation}. 
To limit the number of free parameters, we incorporate all these factors in the two isomerization probabilities 
$P_{\mathrm{CP}}(\mathit{t}\rightarrow \mathit{c})$ and $P_{\mathrm{CP}}(\mathit{c}\rightarrow \mathit{t})$ for  
\textit{trans}-to-\textit{cis}
and  
\textit{cis}-to-\textit{trans}
isomerization, respectively. When birefrigence has 
relaxed towards zero, 
a steady state is reached, where the numbers of \textit{cis} and \text{trans} isomers fulfill 
$N_c P_{\mathrm{CP}}(\mathit{c}\rightarrow \mathit{t}) 
=  N_t P_{\mathrm{CP}}(\mathit{t}\rightarrow \mathit{c})$.
Hence, the ratio
\begin{equation}
R=\frac{ P_{\mathrm{CP}}(\mathit{t}\rightarrow \mathit{c}) }{ P_{\mathrm{CP}}(\mathit{c}\rightarrow \mathit{t}) }
= \frac{N_c}{N_t}  \, .
\label{eq: isomerization ratio}
\end{equation}
is an essential parameter of our model.
Previous measurements show that the absorbance of the two isomers is very similar at 514 nm, 
therefore $R$ should be in the order of unity \cite{yi2011dynamics}.
Because light-induced isomerization cycles occur at a much faster rate than the spontaneous \textit{cis}-\textit{trans} relaxation, the latter is neglected during CP erasure. 
Importantly, we assume that  after an isomerization event,
the orientation of the molecule is chosen at random \cite{yi2008high,kiselev2009kinetics}.
 
As anticipated in Sec.~\ref{sec: experiment}, because of the high in-plane packing density of the SAM, 
molecules must overlap during the relaxation process in order to become randomly oriented. 
Following again a minimal approach, and because we restricted molecular motion to a plane, 
we allow the molecules to overlap by introducing a simple interaction potential $U(\mathcal{O}) =U_{\mathrm{step}} \mathcal{O}$,
which is proportional to the number of overlaps $\mathcal{O}$ 
and the energetic cost
$U_{\mathrm{step}}$
of each overlap, which sets the unit of energy.
We clarify below how $U(\mathcal{O})$ affects the rotational dynamics of our model molecules.

\subsection{\label{sec: kinetic monte carlo} Kinetic Monte Carlo simulation}

The system dynamics is generated using a kinetic Monte Carlo algorithm \cite{battaile2008kinetic}.
Rotational 
dynamics is implemented
by picking a molecule at random and  rotating it by an angle $\delta \theta$
chosen with equal probability from 
the interval $[-\Theta,\Theta]$. The maximum 
rotational step size $\Theta$ is connected to the 
molecular self-diffusion constant $D_{\theta}$ via a Monte Carlo time step $\mathit{dt}$ \cite{patti2012brownian}
\begin{equation}
 \Theta = \sqrt{2 D_{\theta} \mathit{dt}} \, .
\end{equation}
The Monte Carlo time step $\mathit{dt}$ is set such that a single Monte Carlo trial move is accepted with a rate
close to one, which avoids non-local moves and guarantees a reliable dynamics \cite{battaile2008kinetic,sanz2010dynamic,patti2012brownian}. 
In Sec.~\ref{sec: needles relaxation} we fix $\mathit{dt}= 10^{-4}$ while in Sec.~\ref{sec: relaxation isomerization} 
we fix $\mathit{dt}=10^{-5}$ to increase the time resolution. 
Both values ensure an acceptance probability close to one. 
For the self-diffusion constant of our 
simple molecular model, we 
rely on
%\pc{\st{use that of} get inspired by 
the result for a very long cylinder  
\cite{lowen1994brownian,kahlitz2012clustering}
\begin{equation}
D_{\theta} = \frac{3D_0}{\pi L^2}\big( \ln \sigma - 0.622 + 0.917 / \sigma -0.050 / \sigma^2 ) \, ,
\end{equation}
where $L$ is the cylinder length, $\sigma$ the cylinder aspect 
ratio and $D_0 = k_{\mathrm{B}}T/ (\eta_S L)$ with $k_{\mathrm{}B}$ the Boltzmann 
constant, $T$ the temperature, and $\eta_S$ the shear viscosity of the fluid. 
Since
our needles are infinitely thin, we choose $\sigma = 1000$ and for simplicity, the rotational diffusion constants 
of \textit{trans} and \textit{cis} molecule is assumed to be the same.

Both rotational motion and isomerization take place under the influence of the interaction potential 
$U(\mathcal{O})$. At each Monte Carlo step a molecule is picked at random and rotated and isomerized using the set 
of rules defined in the previous section. After a trial move of a single molecule, the number of overlaps and the 
energy $U(\mathcal{O}_{\mathrm{new}})$ of the new configuration are evaluated and compared to 
the
old configuration with $U(\mathcal{O}_{\mathrm{old}})$. Following the
standard Metropolis scheme, the move is accepted with probability
\begin{equation}
p=\min {1,\exp(-1/(k_{\mathrm{B}}T)[U(\mathcal{O}_{\mathrm{new}})-U(\mathcal{O}_{\mathrm{old}})])} \, .
\end{equation} 
In the following we express $k_{\mathrm{B}}T$ in units of $U_{\mathrm{step}}$.

A complete sweep consists of $N$ trial moves and the 
running Monte Carlo time 
is measured in units of $dt$. 
Thus, $t_{MC} = N_s$, where $N_s$ is the number of 
Monte Carlo sweeps. 
The number density
of the model molecules
is defined as $\rho=N/l^2$, where $l$,
in units of $L$, 
is the side length of the 
square simulation box. 
For all the results presented in the following, we use a total of $N=10000$ molecules 
under periodic boundary conditions.

As initial condition, we use a configuration, in which both isomers are equally present.
The \textit{trans} molecules exhibit orientational order while \textit{cis} molecules are  
randomly oriented. 
Starting from this configuration, we then follow the relaxation of the birefringence towards equilibrium.

%%
%% Birefringence Relaxation
%%

\subsection{Birefringence Relaxation}

\begin{figure}
\begin{center}
\includegraphics{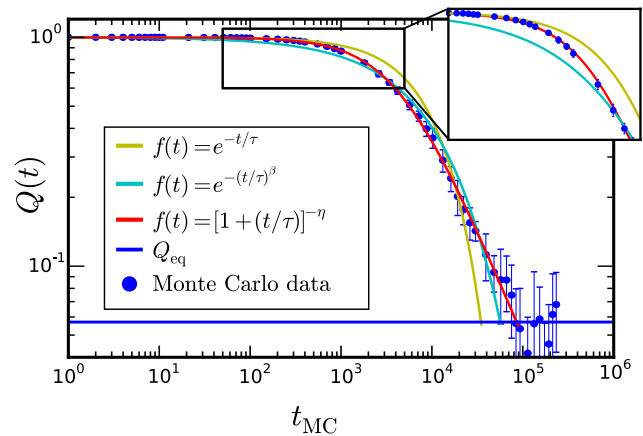}
\caption{(colors) Time evolution of $Q(t)$ as defined in Eq.~(\ref{eq: birefringe model}). Blue points are the kinetic Monte Carlo data. 
Fitting functions are shown as continuous lines.
The horizontal blue line gives the equilibrium value $Q_{\mathrm{eq}}$ of $Q(t)$.
The inset 
zooms in on
the squared area with log scale on the horizontal axis and linear scale on the vertical axis.
The simulation data correspond to a thermal erasure process with $\Temp=4.5$, $\rho=20$, $S(0)=0.633$, $N_{\mathrm{t}}(0)=N$ and $\mathit{dt}=10^{-4}$. }
\label{fig: relaxation fit}
\end{center}
\end{figure}

The degree of alignment within the system is evaluated by the nematic order parameter $S(t)$. 
Since
\textit{trans} isomers have much higher shape anisotropy than \textit{cis} isomers \cite{pedersen1998quantum}, 
we evaluate the nematic order parameter only for the \textit{trans} isomers.
We calculate the nematic order parameter as the positive eigenvalue of the tensor order parameter
\begin{equation}
T_{\alpha \beta } (t) = \langle 
N^{-1}_{\mathit{t}}(t)\sum_{\substack{i=1 \\ i \in \left\{ \mathit{trans} \right\} }}^{N_{\mathit{t}}(t)} (2u_{\alpha}^{i}(t)u_{\beta}^{i}(t)-\delta_{\alpha \beta})
\rangle \, ,
\end{equation}    
where $u_{\alpha}^{i}(t)$ is the $\alpha$-th Cartesian coordinate of the unit vector 
pointing along the central segment of the $i$-th molecule in the \textit{trans} configuration 
at time $t$ and $\langle \ldots \rangle$ denotes non-equilibrium averaging,
for which we used at least 10 different runs.
For a perfectly aligned system $S=1$. The monolayer birefringence $\Delta n(t)$ is proportional 
to both the degree of molecular alignment and the number of rod-like molecules \cite{elston1998optics}. 
Therefore, to monitor 
birefringence relaxation, we keep track of
\begin{equation}
Q(t)=\frac{\Delta n (t)}{\Delta n(0)} = \frac{S(t)N_{\mathit{t}}(t)}{S(0)N_{\mathit{t}}(0)} \, .
\label{eq: birefringe model}
\end{equation}
The data points in Fig.~\ref{fig: relaxation fit} give a typical temporal relaxation of $Q(t)$ from our kinetic Monte Carlo scheme.

We now illustrate how we discriminate between three possible functional forms of the relaxation dynamics of $Q(t)$. 
Recall that simple relaxation processes are expected to show an exponential form $\phi(t)=e^{-t/\tau}$ 
(a Maxwell-Debye relaxation), which is
characterized by a well-defined time $\tau$ that fully determines the kinetics of the system. 
In some systems, however \cite{shlesinger1988fractal}, the relaxation significantly deviates from an exponential form  and is
described either by a stretched exponential decay $\phi(t)=e^{-(t/\tau)^\beta}$ with $0<\beta<1$ or by 
an asymptotic power law as in Eq.~(\ref{eq: PSSG relaxation}).

The following procedure is used to discriminate between an exponential, a stretched-exponential, 
and an asymptotic power-law relaxation of $Q(t)$. For each choice of model parameters, 
we run kinetic Monte Carlo simulations until $Q(t)$ has reached a clear steady-state 
value $Q_{\mathrm{eq}}$ (illustrated as an horizontal line in Fig.~\ref{fig: relaxation fit}). 
%\pc{\st{$Q(t)$ is averaged over 10 independent runs.}} 
The equilibration time $t_{\mathrm{eq}}$ is the first time for which $Q(t_{\mathrm{eq}})=Q_{\mathrm{eq}}$. 
$Q(t)$ is then fitted by least-square minimization over the range $t \in [0,t_{\mathrm{eq}}]$ with the three functional forms
given in Fig.\ \ref{fig: relaxation fit}. 

For each fit curve a goodness-of-fit test\cite{bevington2003data} is performed. The most reliable fit
function is chosen as the one with the value of the reduced $\chi ^2$ closest to $1$.
In the example given in Fig.~\ref{fig: relaxation fit}, we obtain $\chi ^2 = 9.65$ for the exponential function, $\chi ^2 = 2.32$ for the 
stretched-exponential, and $\chi ^2 = 0.54$ for the asymptotic power law. The power law thus clearly
provides the best fit of the simulation data.  
The results of the fitting procedure in different regions of  parameter space and 
under different initial conditions will be discussed in the  following two sections.

% % % % % % % % % % % % % % % % % % % % % % % % % % % % % % % % %
% % % % % % % % % % % % % % % % % % % % % % % % % % % % % % % % %
% %
% % Section : RESULTS
% %
% % % % % % % % % % % % % % % % % % % % % % % % % % % % % % % % %
% % % % % % % % % % % % % % % % % % % % % % % % % % % % % % % % %

\section{\label{sec: needles relaxation} Results: pure \textit{trans} system}
In this section we present the results of the kinetic Monte Carlo simulations
for a system that only contains \textit{trans} molecules.

\begin{figure}[t]
\begin{center}
\includegraphics{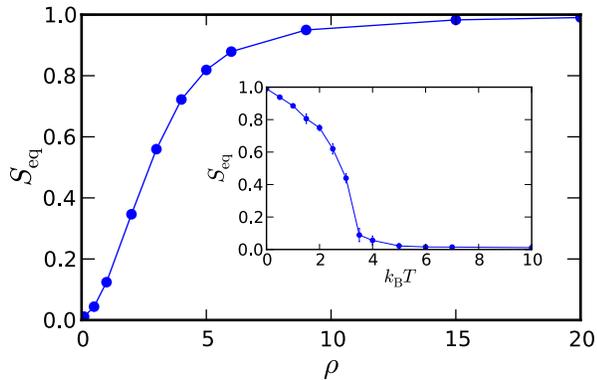}
\caption{(colors). 
%\pc{Finite-size} 
Equilibrium value of the nematic order parameter $S_{\mathrm{eq}}$ in a system of 
\textit{trans} molecules
with finite size.
Main plot: $S_{\mathrm{eq}}$ versus $\rho$ for hard needles that are not allowed to overlap. 
Inset:
$S_{\mathrm{eq}}$
versus $\Temp$ for
fixed $\rho=20$.
Lines are guides for the eye. 
}
\label{fig: pure needle nematic}
\end{center}
\end{figure}

First, we characterize the equilibrium properties of the system with temperature $T$ and density $\rho$.
Figure \ref{fig: pure needle nematic} shows the equilibrium value of the nematic order parameter, $S_{\mathrm{eq}}$, determined after 
equilibration. The main plot refers to molecules with hard-core interactions that cannot overlap at all.
The steady-state value of the nematic order parameter shows a steep increase with $\rho$.
The inset in Fig.~\ref{fig: pure needle nematic} instead plots $S_{\mathrm{eq}}$ versus $\Temp$ at $\rho=20$.
If the temperature is sufficiently high, molecules can pass over their neighbors by creating overlaps and thereby drastically
reduce $S_{\mathrm{eq}}$. So, allowing the molecules to overlap, results in an isotropic state at sufficiently high 
temperatures even at high in-plane packing density.

\begin{figure}[t]
\begin{center}
\includegraphics{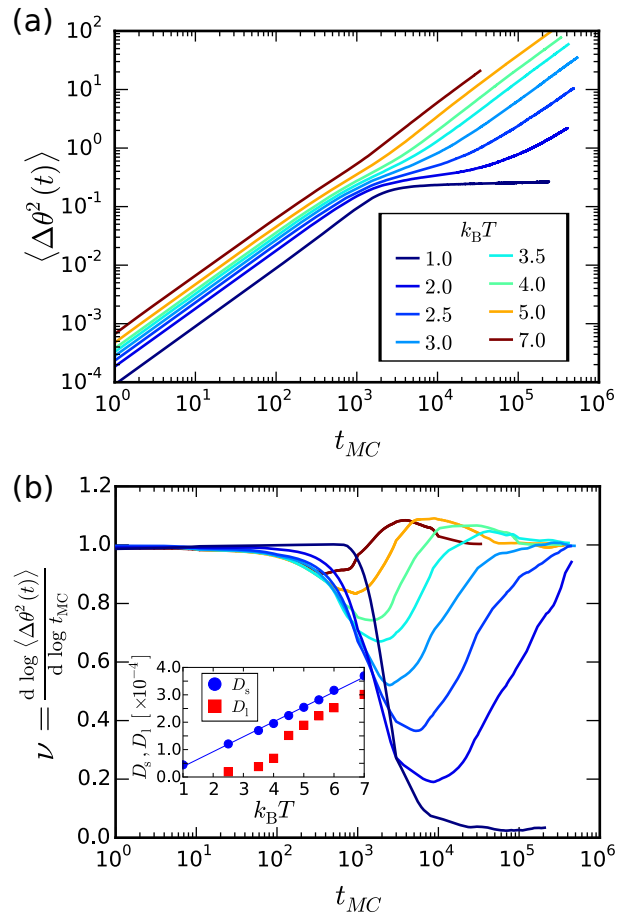}
\caption{(colors). Rotational dynamics in a system of 
\textit{trans} molecules 
for different $\Temp$,
at density $\rho=20$,
and with initial nematic order 
$S(0)=0.6$. The legend in (a) applies to both  
graphs.
(a) 
Rotational mean square displacement $\langle \Delta \theta ^2 (t) \rangle$
versus time. 
(b) The logarithmic derivative of $\langle \Delta \theta ^2 (t) \rangle$ to extract the local exponent $\nu$
in $\langle \Delta \theta ^2 (t) \rangle \propto t^{\nu}$.
The extent of the subdiffusive regime ($\nu < 1$) increases 
with decreasing temperature.
The inset shows the 
short- (blue circles) and 
long-time (red squares) diffusion constants as a function of $\Temp$.
}
\label{fig: needles msd}
\end{center}
\end{figure}

We now characterize the relaxation dynamics of the birefringence at different temperatures. 
We fix the initial degree of the nematic order to $S(0)=0.6$ 
(the same qualitative results are obtained using values from $S(0)\approx 0.5$ to $S(0) \approx 1.0$)  
and follow the temporal evolution of $S(t)$ while it relaxes back to its equilibrium value $S_{\mathrm{eq}}$.
First, we analyze the rotational diffusion of the molecules,
which results in the decay of $Q(t)$, by looking at the rotational mean square displacement  
\begin{equation}
\langle \Delta \theta ^2 (t) \rangle = \big \langle \frac{1}{N} \sum _{i=1}^{N} ( \theta_i (t) - \theta_i (0) )^2 \big \rangle \, ,
\label{eq: rotational msd}
\end{equation} 
where
$\theta_i(t)$ 
is the orientation angle
of the $i$-th molecule at time $t$. 
In Fig.~\ref{fig: needles msd}(a), we plot  
$\langle \Delta \theta ^2 (t) \rangle$ 
versus $t_{\mathrm{MC}}$ for different temperatures in a system with fixed density $\rho=20$.
In addition,
Fig. \ref{fig: needles msd}(b) shows the 
logarithmic
derivative of the rotational mean square displacement, which gives the local exponent $\nu$
in
$\langle \Delta \theta ^2 (t) \rangle \propto t^{\nu}$. 
Anomalous diffusion  has $\nu \ne 1$.
Initially, the molecules diffuse with a diffusion constant $D_s$ that increases linearly 
with temperature,
as expected from the Einstein relation [see inset of Fig. \ref{fig: needles msd}(b)]. 
When the needles start to overlap,  
a subdiffusive regime emerges and its extent increases with
decreasing temperature [see Fig.\ \ref{fig: needles msd}(b)]. 
Ultimately,
normal diffusion is recovered 
even in systems
where nematic order is well developed (for instance at $\Temp=2.5$ one finds $S_{\mathrm{eq}}\approx 0.6$ in
Fig.~\ref{fig: pure needle nematic}).
In order to diffuse, molecules have to pass each other by creating overlaps. 
At low temperatures, 
crossing energy barriers makes rotational diffusion an activated process.
Hence, the 
long-time diffusion constant $D_l$ 
is no longer linear in temperature [see inset of 
Fig.~\ref{fig: needles msd}(b)].

\begin{figure}[thc]
\begin{center}
\includegraphics{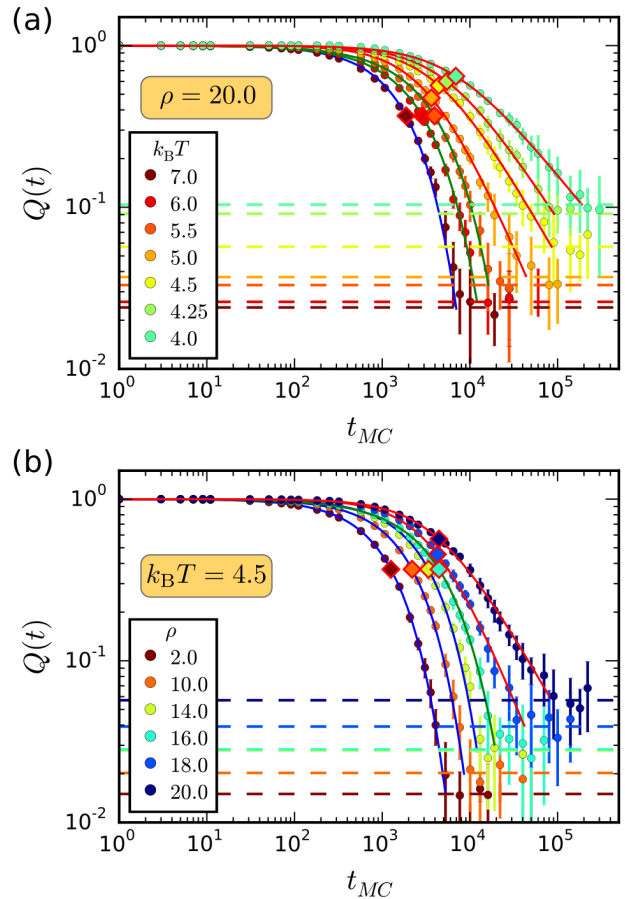}
\caption{(colors). Relaxation of the birefringence $Q(t)$ in a system of  \textit{trans} molecules with initial  
nematic order $S(0)=0.6$: (a) for different $\Temp$ at density $\rho=20$, 
and 
(b) 
%\pc{and} 
at different $\rho$ for $\Temp=4.5$. Circles show the numerical results and the continuous lines the
best-fitting functions with blue for exponential, green for stretched-exponential, and red for power-law decay. 
The characteristic times of the relaxation curves are indicated as diamonds. 
Horizontal dashed lines show the equilibrium value of $Q(t)$ with matching colors.
}
\label{fig: needles birefringence}
\end{center}
\end{figure}
 
Rotational
diffusion 
causes initial nematic order to fully decay for $\Temp \ge 4$.
Figure~\ref{fig: needles birefringence}(a) shows the relaxation
of $Q(t) = S(t) / S(0)$ 
at fixed $\rho=20$ and different $\Temp$, while in Figure~\ref{fig: needles birefringence}(b) we keep temperature at
$\Temp=4.5$ and vary density $\rho$. In a high-temperature or low-density regime the relaxation is very well fitted by an exponential function [$\Temp \ge 7.0$ in Fig.~\ref{fig: needles birefringence}(a) and $\rho \le 14.0$ in Fig.~\ref{fig: needles birefringence}(b)], while in a high-density, low-temperature regime [$\Temp \le 5.0$ in Fig.~\ref{fig: needles birefringence}(a) and $\rho \ge 18.0$ in Fig.~\ref{fig: needles birefringence}(b)] the 
power-law decay defined in Eq.~(\ref{eq: PSSG relaxation}) provides an excellent fit of the simulation data. 
In this regime
both the exponential and stretched-exponential functions give significant residual errors. 
Our Monte Carlo results are best fitted by a  stretched exponential function  
with $\beta \approx 0.8$ in an intermediate regime [$\Temp=5.5,6$ in Fig.~\ref{fig: needles birefringence}(a) and $\rho=16.0$ in 
Fig.\ \ref{fig: needles birefringence}(b)]. 
%The 
To summarize, the
interaction with neighboring molecules, which is more 
relevant at low temperatures and high densities, causes a transition from an exponential decay of
$Q(t)$ to a power-law relaxation.

\begin{figure}[t]
\begin{center}
\includegraphics{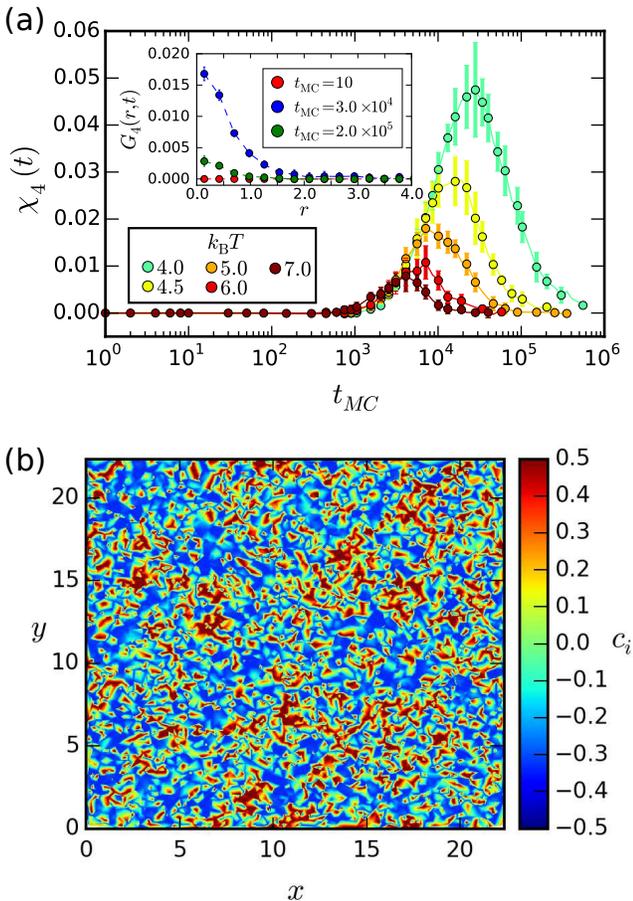}
\caption{(colors). (a) Temporal evolution of the dynamical susceptibility $\chi_4(t)$  
for different $\Temp$
at $\rho=20$ 
and with
$S(0)=0.6$. 
Inset:
Four-point correlation function $G_4(r,t)$
plotted versus $r$ at different times for
$\Temp=4.0$ and $\rho=20.0$.
The peak in $\chi_4(t)$ 
indicates extended domains of correlated rotational motion. 
(b) The 
mobility
$c_i(t)$ defined in Eq.~(\ref{eq: mobility}) 
color-coded in
the whole simulation box for 
$\Temp=4.0$, $\rho=20.0$,
and
at time $t_{\mathrm{MC}}=3.0 \times 10^{4}$, 
right at
the maximum of $\chi_4 (t)$.
}
\label{fig: needles susceptibility}
\end{center}
\end{figure}

As anticipated in Sec.~\ref{sec: Model and Method}, non-exponential  
relaxation originates
from the presence of a 
wide
distribution of 
relaxation times
in the system. 
In the concept of dynamic heterogeneity \cite{berthier2011phys,berthier2011theoretical}, 
this distribution is traced back to spatial and temporal variations
in the local structure of the  system, which then determines its dynamic evolution.
Molecules diffusing slower or faster than the average  
become spatially correlated, giving rise to regions with slow and fast dynamics. Hence, averaging the dynamics over this heterogeneous  
environment
leads to an overall non-exponential relaxation. 
A typical quantity to monitor this dynamic heterogeneity is a four-point correlation function, which we 
introduce here for the angular displacement following Ref.~[\onlinecite{berthier2011phys}].
We define the mobility $c_i(t)$, in order to quantify how mobile the molecule $i$ is,
\begin{equation}
c_i(t) = \exp [-\Delta \theta_i (t)^2 ] - \frac{1}{N} \sum_{j=1}^{N} \exp [-\Delta \theta_j (t)^2 ]   \, ,
\label{eq: mobility}
\end{equation}     
where $\Delta \theta_i (t)^2$ is defined as in Eq.~(\ref{eq: rotational msd}). The variable $c_i(t)$ has zero mean and 
is positive (negative) if the $i$-th molecule 
moves
less (more) than the average. 
The four-point correlation function is then
\begin{equation}
G_4(r,t)=\frac{\sum_{i,j} c_i(t)c_j(t)\delta(r-\vert \vec{r}_{ij} \vert) }{\sum_{i,j} \delta(r-\vert \vec{r}_{ij} \vert)} \, ,
\label{eq: four-point correlation}
\end{equation}    
where $\vert \vec{r}_{ij} \vert$ is the distance between molecule $i$ and $j$ and $\delta$ is the Dirac delta function. 
The correlation function $G_4(r,t)$ measures the spatiotemporal correlations in the dynamics of the molecules 
over a distance $r$ at time $t$. The presence of correlated domains in space and the degree of dynamic heterogeneity
is monitored by the dynamical four-point susceptibility
for the angular displacement
\cite{berthier2011theoretical,franz2000non},
\begin{equation}
\chi _4(t)=\int \mathrm{d}r \, G_4(r,t) \, .
\label{eq: chi four}
\end{equation}
It increases with 
%the 
increasing size of correlated domains.

Figure~\ref{fig: needles susceptibility}(b) 
illustrates the heterogeneity in the rotational mobility of the molecules by plotting
the color-coded  
$c_i(t)$ 
in
the whole simulation box for $\Temp=4.0$ and $\rho=20$. 
Extended domains of correlated molecular rotations are clearly visible.
The inset in Figure~\ref{fig: needles susceptibility}(a) shows the corresponding four-point correlation function $G_4(r,t)$ 
for $\Temp=4.0$
at $t_{MC} = 3.0 \times 10^{4}$, where the correlations in time and space are largest compared to other times.
In order to monitor the complete temporal evolution of the dynamical heterogeneities,
we present the susceptibility $\chi _4(t)$ in
the main graph of 
Fig.~\ref{fig: needles susceptibility}(a) 
for different 
temperatures 
and at density $\rho=20$.
The dynamical susceptibility shows a clear peak, 
which coincides with the time window, during which the power-law relaxation of $Q(t)$ is observed
in Fig.~\ref{fig: needles birefringence}(a).
The color plot of Fig.\ \ref{fig: needles susceptibility}(b) 
is obtained at the maximum of $\chi _4(t)$ at $\Temp=4$.
The maximum increases with decreasing temperature and therefore demonstrates that
regions of correlated motion become more relevant at low temperatures.

\begin{figure}
\begin{center}
\includegraphics{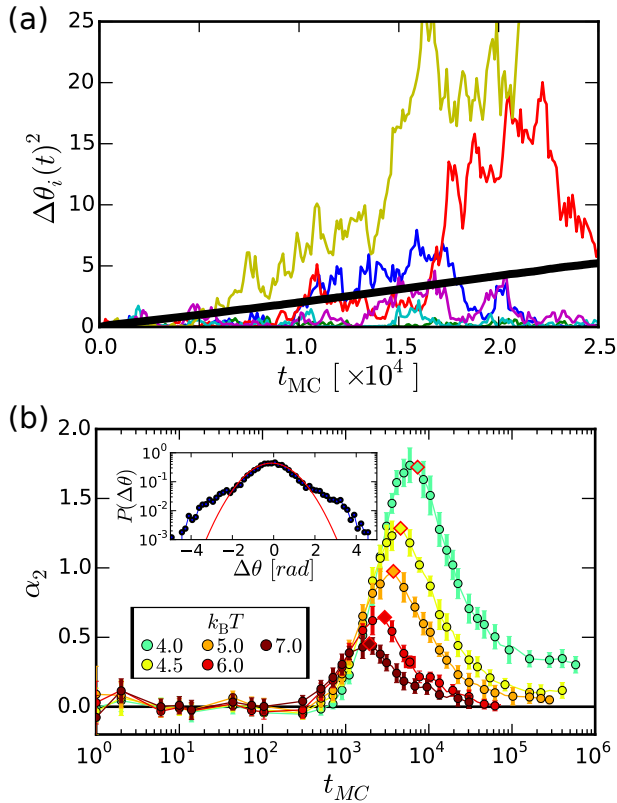}
\caption{(colors). (a) 
Square
displacement of selected individual molecules plotted versus time for  
$\Temp=4.0$ and $\rho=20.0$. The thick black line  
is the mean value over all molecules. 
(b) Temporal evolution of the non-Gaussian parameter $\alpha_2$ of  
the probability distribution $P(\Delta \theta (t))$ 
for different $\Temp$
at $\rho=20$
and with
$S(0)=0.6$.
Values of $\alpha_2 \ne 0$ 
indicate non-Gaussian tails in the distribution $P(\Delta \theta (t))$ of rotational steps, i.e.,
an excessive number of
molecules significantly faster or slower than the average. 
Inset: Example of $P(\Delta \theta (t))$ at $t_{MC} = 7.0 \times 10^{3}$ and for $\Temp = 4.0$. The red line shows a Gaussian 
distribution.
}
\label{fig: needles kurtosis}
\end{center}
\end{figure}

Spatio-temporal variations in the local environment of the pinned molecules are also responsible for the anomalous diffusion,
which we demonstrate
in Fig.\ \ref{fig: needles msd}. In Fig.\ \ref{fig: needles kurtosis}(a) we show individual molecular trajectories for $\Temp=4.0$ and 
$\rho=20.0$. One already senses that some molecules rotate significantly faster than the average (shown as a thick black line), 
while all trajectories indicate that rotational diffusion proceeds by sudden jumps.
This feature, together with the spatio-temporal correlations discussed before, are common signatures of a glass-like dynamics
\cite{kawasaki2011structural,zheng2011glass,hurley1995kinetic,teboul2013isomerization} . 
In particular, it has been 
suggested
recently 
that a non-Gaussian distribution of molecular displacements
is a
universal feature of finite-dimensional glass-like dynamics \cite{chaudhuri2007universal}.     
To investigate this point, we evaluated the non-Gaussian parameter $\alpha_2$ of
the distribution of molecular displacement $P(\Delta \theta)$.
In two dimensions it is calculated using the second and fourth moment of  $P(\Delta \theta)$,
\begin{equation}
\alpha_2 (t) = \frac{\langle \Delta \theta ^4 (t) \rangle}{\langle \Delta \theta ^2 (t) \rangle^2}-3 \, ,
\end{equation}
where for a Gaussian distribution 
one has $\alpha_2=0$. In Fig.\ \ref{fig: needles kurtosis}(b), 
we show
$\alpha_2$  
for different temperatures $\Temp$ and at density  $\rho=20$.
The distribution $P(\Delta \theta)$ becomes highly non-Gaussian as also demonstrated in the inset, which shows 
$P(\Delta \theta)$ for $\Temp=4.0$ and at the time,
when
$\alpha_2(t)$
maximal.
Fast rotating molecules are responsible for the non-Gaussian tails of the distribution. The
peak in $\alpha_2$ increases and shifts to later times upon cooling, which again coincides
with the developing non-exponential relaxation of the birefringence.
Interestingly, the characteristic times  
in the relaxation laws for $Q(t)$, shown  
as diamond markers in Fig.~\ref{fig: needles kurtosis}(a),
agree nicely with the locations of the maxima in $\alpha_2(t)$.
           
% % % % % % % % % % % % % % % % % % % % % % % % % % % % % % % % % %
% % % % % % % % % % % % % % % % % % % % % % % % % % % % % % % % % %
% RESULTS WITH ISOMERIZATION: thermal erasure
% % % % % % % % % % % % % % % % % % % % % % % % % % % % % % % % % %
% % % % % % % % % % % % % % % % % % % % % % % % % % % % % % % % % %

\section{\label{sec: relaxation isomerization} Results: Relaxation with isomerization}

In this section we 
explore the relaxation of birefringence $Q(t)$ in a system, in which both \textit{trans} and \textit{cis} molecules are present.
In particular, we
investigate how the 
nearly isotropic \textit{cis} molecules
influence the relaxational dynamics of $Q(t)$
under two different isomerization scenarios: thermal erasure and CP erasure.

The model has a  rather rich parameter space and its full exploration is beyond the scope of this paper.  
In the following, we
fix the initial conditions to be close to the experimental values and explore
the relaxational dynamics of $Q(t)$ for 
different isomerization rates.
We
prepare the initial state 
with an equal number of  \textit{trans} and \textit{cis} molecules,
$N_t(0)=N_c(0)$, and 
order parameter
$S(0)=0.6$; both values are close to the estimates carried out in Ref.~[\onlinecite{fang2013athermal}].
The
\textit{cis} molecules 
are 
randomly oriented at $t=0$ 
and do not contribute to $S(t)$.
In this section we  
choose the time step
$\mathit{dt}=10^{-5}$ to increase the temporal resolution, which 
results in an
equilibration time of up to 9 weeks on an Intel Xeon X5550 machine with a 2.66 GHz CPU. 
In
order to 
compare
the results of this section directly 
with the results of Sec.~\ref{sec: needles relaxation}, we still give
the Monte-Carlo time
$t_{\mathrm{MC}}$ in units of $\mathit{dt}=10^{-4}$.

\begin{figure*}[thbc]
\begin{center}
\includegraphics{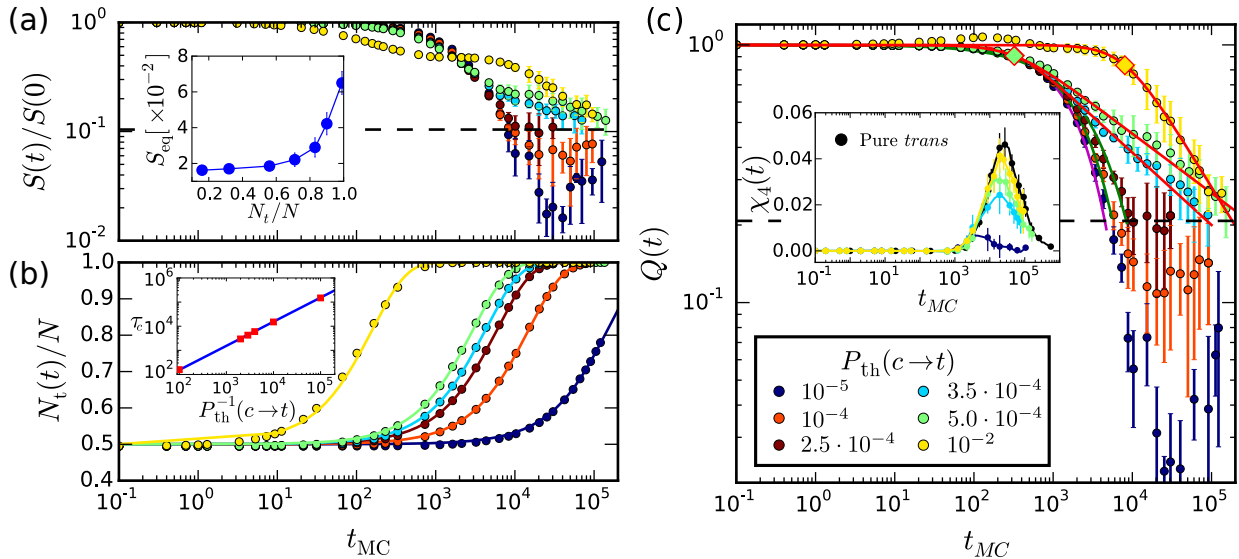}
\caption{(colors). Relaxation towards equilibrium during thermal erasure in a system with $\rho=20.0$ and $\Temp=4.0$. 
The legend
in (c) applies to all 
graphs.
(a) Temporal evolution of the nematic order parameter $S(t)/S(0)$. 
The horizontal dashed line marks the equilibrium value to which all the curves 
have to
converge. The inset 
shows
the equilibrium value 
$S_{\mathrm{eq}}$
as a function of the relative number of \textit{trans} isomers, 
$N_t/N$. 
(b) Relative number of \textit{trans} isomers, 
$N_t/N$,
during thermal erasure. Fits of the 
simulation data
with Eq.~(\ref{eq: trans relaxation}) are shown as 
solid
lines with matching colors. The inset 
plots the isomerization time
$\tau_{\mathrm{c}}$ versus
$P^{-1}_{\mathrm{th}}(c \rightarrow t)$
(red squares). The 
fitted
blue line is $\tau_{\mathrm{c}}=1.54 \, P^{-1}_{\mathrm{th}}(c \rightarrow t)$. 
(c) Temporal relaxation of the birefringence $Q(t)$.  
The horizontal dashed line marks the equilibrium value of $Q(t)$. 
The full 
lines are the best fits to an exponential (magenta), a stretched exponential (green), or a power law (red).
Diamonds indicate the
characteristic times
of the power law relaxation.
The inset shows the dynamical susceptibility $\chi_4 (t)$
for \emph{trans} molecules already present at $t=0$.
As a reference
the black circles give $\chi_4(t)$ for the pure \textit{trans} system at 
the same $\rho$ and $\Temp$.
}
\label{fig: thermal erasure summary}
\end{center}
\end{figure*}

\subsection{\label{subsec: thermal relaxation} Thermal erasure}

Here, we
discuss the relaxation of the birefringence,
when the 
monolayer is not illuminated.
Since
the \textit{trans} isomer is the ground-state of the dMR, 
all the molecules in 
the
\textit{cis} configuration will isomerize back to 
the
\textit{trans} 
state
after some characteristic time. As discussed in detail in 
Sec.~\ref{sec: molecular model}, the 
\textit{cis} to \textit{trans} isomerization rate
of isolated molecules is
the isomerization probability
$P_{\mathrm{th}}(\mathit{c}\rightarrow \mathit{t})$. 
  
In view of the results discussed in 
Sec.\ \ref{sec: needles relaxation},
we expect the relaxation
dynamics
to be exponential in the high-temperature
and
low-density regime,
regardless of
the isomerization probability. Therefore,
we fix both temperature and density to 
$\Temp=4.0$ and $\rho=20.0$, 
respectively,
where the pure \textit{trans} system shows a clear power-law decay of the birefringence, and monitor how this
decay
is influenced by the isomerization 
rate.

Figure \ref{fig: thermal erasure summary}(b) shows the
temporal evolution of the number of \textit{trans} isomers
for different $P_{\mathrm{th}}(\mathit{c}\rightarrow \mathit{t})$.
We expect an exponential relaxation and, indeed, the
Monte Carlo data 
(circles)
are 
well
fitted with
\begin{equation}
\frac{N_{\mathrm{t}}(t)}{N}=1-\frac{N_{\mathrm{c}}(0)}{N}e^{-t/\tau_{\mathrm{c}}} \, ,
\label{eq: trans relaxation}
\end{equation}
where 
the fit parameter, the relaxation time $\tau_{\mathrm{c}}$, is proportional to the inverse isomerization probability,
$\tau_{\mathrm{c}} = 1.54 \, P_{\mathrm{th}}^{-1}(c \rightarrow t)$ 
[see inset of Fig.~\ref{fig: thermal erasure summary}(b)]. 
We find
$\tau_{\mathrm{c}} > P_{\mathrm{th}}^{-1}(c \rightarrow t)$ 
because
in a crowded environment 
some 
of the attempted isomerization events are rejected as they generate more overlaps between the molecules.

\begin{figure*}
\begin{center}
\includegraphics{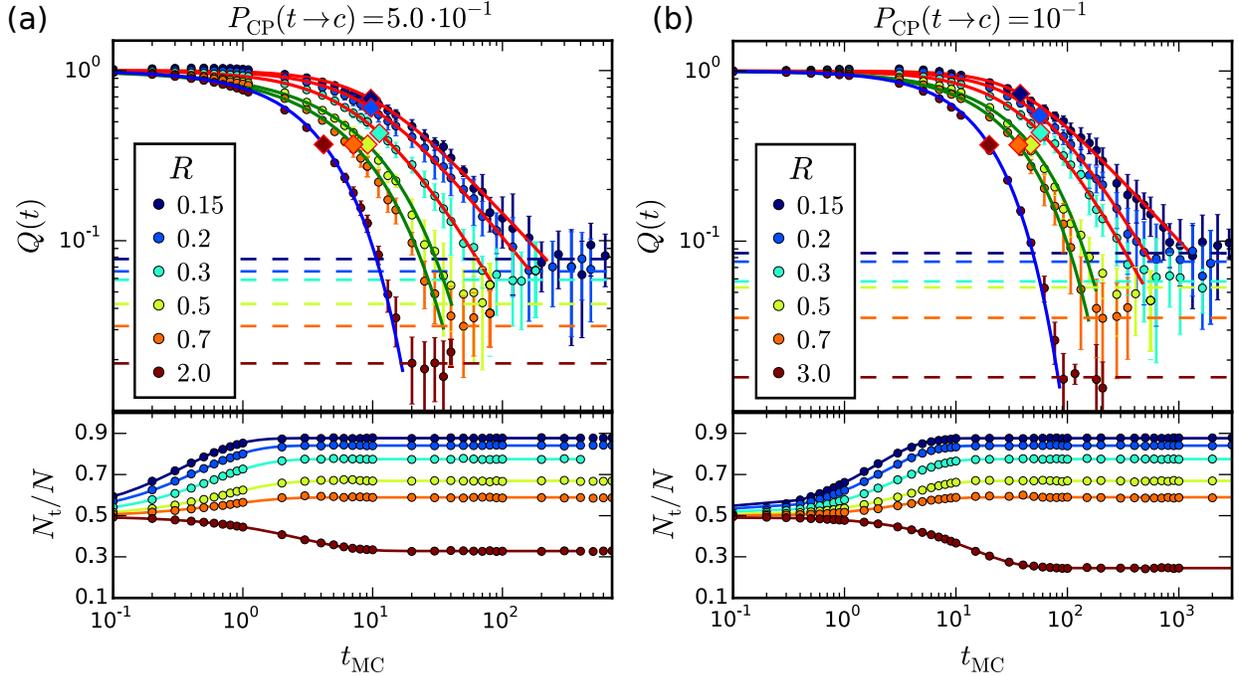}
\caption{(colors). Relaxation of the birefringence $Q(t)$ and of the 
relative number of
\textit{trans} isomers, 
$N_t(t)/N$, 
during CP erasure 
for different ratio $R$
at fixed $\rho=20$ and $\Temp=4$.
The \textit{trans}-to-\textit{cis} isomerization rate is $P_{\mathrm{CP}}(\mathit{t}\rightarrow \mathit{c}) = 5.0 \cdot 10^{-1} $ in (a) 
and $P_{\mathrm{CP}}(\mathit{t}\rightarrow \mathit{c}) = 10^{-1}$ in (b).
The \textit{cis}-to-\textit{trans} isomerization rates are 
$P_{\mathrm{CP}}(\mathit{c}\rightarrow \mathit{t}) = P_{\mathrm{CP}}(\mathit{t}\rightarrow \mathit{c}) / R$.
The best-fitting function for the relaxation of $Q(t)$ are shown as  
solid
lines, blue for the exponential, green for the stretched-exponential, and red for the power-law relaxation  
of
Eq.~(\ref{eq: PSSG relaxation}). 
The diamond markers with matching colors indicate the characteristic times $\tau_t$.
The fitted exponents
are given in Table\ \ref{tab: exponent}. The horizontal dashed lines mark the 
the steady-state value
of $Q(t)$ for different 
$R$ with matching colors. The 
numerical results for $N_t(t)/N$ are 
fitted with Eq.\ \ref{eq: trans number fit CP erasure}  
(solid lines).
}
\label{fig: CP erasure summary}
\end{center}
\end{figure*}

In Figs.~\ref{fig: thermal erasure summary}(a) and (c) we show the respective
temporal evolutions of the nematic order parameter $S(t)/S(0)$ and 
the birefringence $Q(t) \propto N_t(t)S(t)$,
which originate
from the alignment of the \textit{trans} isomers.
We also evaluate the dynamical susceptibility $\chi_4(t)$ defined in Eq.~(\ref{eq: chi four}), but only on the subset of molecules that are in the \textit{trans} 
configuration at the beginning of the simulation
at  
$t_{\mathrm{MC}}=0$. The results are plotted in the inset of Fig.~\ref{fig: thermal erasure summary}(c), where we also 
include
the dynamical susceptibility for the pure \textit{trans} system at $\rho=20$ at $\Temp=4.0$ (black circles) 
as
a reference.
This 
will
allow us to 
to quantify how
the presence of the \textit{cis} isomers
influences
the development of 
dynamic heterogeneities.
The 
temporal
relaxation 
of both $S(t)$ and $Q(t)$ strongly depends on the isomerization rate. 
Analyzing Figs.\ \ref{fig: CP erasure summary}(a), (b) and (c),
we identify four
different regimes:

\textbf{1)} 
For sufficiently small relaxation rate 
[$P_{\mathrm{th}}(\mathit{c}\rightarrow \mathit{t}) = 10^{-5}$ in Fig.~\ref{fig: thermal erasure summary}], 
the birefringence $Q(t)$ relaxes exponentially.
On the time scale of the declining $S$, the
number of \textit{cis} molecules stays 
constant [compare plots (a) and (b)]. They thus 
create a more uniform environment, 
as indicated by the nearly vanishing dynamical susceptibility $\chi_4$,
in which $S(t)$ and $Q(t)$ 
relax exponentially and much faster than the pure \emph{trans} system for the same temperature and
density in Sec.\ \ref{sec: needles relaxation}.
Note that at late times $Q(t)$ first falls below its equilibrium value (dashed horizontal line) and then increases again.
The reason is that the equilibrium value of $S$ depends on the number of \emph{trans} molecules [inset of 
Fig.~\ref{fig: thermal erasure summary}(a)] because the presence of the nearly isotropic \textit{cis} molecules decreases 
$S_{\mathrm{eq}}$. 
Since  
$N_{\mathrm{t}}$ increases for $t_{MC} > 10^4$, $Q(t)$ also increases.

\textbf{2)}  
For $P_{\mathrm{th}}(\mathit{c}\rightarrow \mathit{t})=10^{-4}$ and $2.5 \cdot 10^{-4}$
the best fitting function for the temporal evolution of $Q(t)$ is provided by a stretched exponential 
with
$\beta \approx 0.8$
indicating the transition to the power-law decay.

\textbf{3)} At intermediate isomerization rates  
$P_{\mathrm{th}}(\mathit{c}\rightarrow \mathit{t})=3.5 \cdot 10^{-4}$ and $5.0 \cdot 10^{-4}$
the power-law relaxation 
provides 
the best
fit of the 
simulation
data. Here, the isomerization of the randomly oriented \textit{cis} molecules into the \textit{trans} 
state
happens 
on the same time scale 
as
the relaxation of the nematic order 
parameter. Thus, $Q(t)$ does not decay in the static disordered distribution of \textit{cis} molecules but in a
dynamic and heterogeneous environment, as demonstrated by the clear peak in $\chi_4(t)$. As a result, the relaxation
of $Q(t)$ follows a power law. 
Its
characteristic time $\tau_t$ 
does not 
change
significantly, but the isomerization probability 
seems
to control the power-law exponent. We find 
$\eta=0.281$ at $P_{\mathrm{th}}(\mathit{c}\rightarrow \mathit{t})=3.5 \cdot 10^{-4}$ and $\eta=0.225$ at 
$P_{\mathrm{th}}(\mathit{c}\rightarrow \mathit{t})=5.0 \cdot 10^{-4}$.

\textbf{4)} The situation changes again if the isomerization 
rate
is very large 
[$P_{\mathrm{th}}(\mathit{c}\rightarrow \mathit{t})=10^{-2} $ in Fig.~\ref{fig: thermal erasure summary}].
All
the \textit{cis} isomers 
rapidly 
isomerize
into \textit{trans}
molecules with random orientation and $S(t)$ drops to $ S(0)/2$.
This is compensated by the 
%The 
resulting increase in $N_t(t)$, 
which 
ultimately
generates a bump in $Q(t)$ at $t_{\mathrm{MC}}\approx 10^2$.
Once 
isomerization 
is completed at $t_{\mathrm{MC}}\approx 10^3$, the system is composed of only \textit{trans} molecules 
the aligment of which relaxes via rotational motion.
This is the same situation as
discussed in Sec.\ \ref{sec: needles relaxation}. 
The environment is heterogeneous as indicated by the peak in the dynamical susceptibility, which is nearly as large as in the
pure \textit{trans} system of Sec.\ \ref{sec: needles relaxation}. Without
the initial bump,
$Q(t)$ 
is
fitted well 
by a power-law
with a larger
characteristic time  
and a larger
power-law exponent 
$\eta \approx 0.5$ 
as compared
to case \textbf{3)} 
but similar to the pure \emph{trans} system.

Summarizing the results in
Fig.~\ref{fig: thermal erasure summary}, 
we find that \textit{cis} isomers pinned at random positions not only accelerate the birefringence relaxation compared to
the pure \textit{trans} system but also prevent the 
%\pc{\st{formation} 
development of the
dynamical 
%\pc{\st{heterogeneities} 
heterogeneity
that
%\pc{\st{are} 
is responsible for a non-exponential decay.
We attribute this behavior to the nearly isotropic shape of \textit{cis} isomers, which, regardless of their orientations,
create 
%\pc{\st{the same} 
a similar
environment for the exisiting \emph{trans} isomers. In contrast, newly formed \emph{trans} isomers can adjust their
orientations to their neighbors and thereby allow for the 
formation 
%growth} 
of dynamical heterogeneities. In the experiments of 
Ref.~[\onlinecite{yi2011dynamics}] the lifetime of \textit{cis} isomers during thermal erasure is estimated as 
$0.7 \mathrm{s}$, while the characteristic time of the power-law decay is measured as 
$\tau_{\mathrm{th}} \approx 2 \mathrm{s}$. These values are achievable by isomerization rates between our cases 
\textbf{3)} and \textbf{4)}.

% % % % %% % % % % % % % % % % % % % % 
%% Cp erasure
% % % % % % % % % % % % % % % % % % % %

\subsection{\label{subsec: CP erasure} CP erasure}

Illumination of the SAM with CP light induces \textit{trans}-\textit{cis} isomerization cycles at a rate much faster than the spontaneous 
relaxation
due to
thermal erasure.
Therefore,
we neglect the thermally induced \textit{cis}-to-\textit{trans} transition when modeling the CP erasure process.
As discussed in Sec.~\ref{sec: molecular model}, in order to 
take into account the different light absorption 
of the two isomers, we introduce 
the
two 
respective
probabilities, $P_{\mathrm{CP}}(\mathit{t}\rightarrow \mathit{c})$ and  $P_{\mathrm{CP}}(\mathit{c}\rightarrow \mathit{t})$, 
for 
\textit{trans}-to-\textit{cis} and \textit{cis}-to-\textit{trans} 
isomerization.
Since 
light is circularly polarized, the isomerization probabilities 
do not depend on
the molecular orientation.  

In order to limit the computational cost, we do not explore the full range of isomerization probabilities. 
Instead, we choose their values
such
that
the characteristic times of 
birefringence
relaxation 
for
thermal and CP erasure 
matches the experimental
observations, where they differ by
approximately
two decades
in time
(see the diamond markers in Fig.~\ref{fig: experiment scheme}).
Because illumination of the monolayer produces a very
negligible amount of heat
\cite{fang2013athermal}, we fix both density and temperature 
at
$\rho=20$ and $\Temp=4$, exactly as during 
thermal erasure 
discussed in
Sec.~\ref{subsec: thermal relaxation}.
 
In Fig.\ \ref{fig: CP erasure summary} we show the relaxation of $Q(t)$ and $N_t(t)$ 
towards their steady-state values starting with $N_t = 0.5$ at $t=0$.
In Fig.\ \ref{fig: CP erasure summary}(a) we 
set
$P_{\mathrm{CP}}(\mathit{t}\rightarrow \mathit{c}) = 5.0 \cdot 10^{-1}$ while in Fig.\ \ref{fig: CP erasure summary}(b) 
$P_{\mathrm{CP}}(\mathit{t}\rightarrow \mathit{c}) = 10^{-1}$. The backward isomerization rates, $P_{\mathrm{CP}}(\mathit{c}\rightarrow \mathit{t})$, are 
chosen by the ratio $R = P_{\mathrm{CP}}(\mathit{t}\rightarrow \mathit{c}) / P_{\mathrm{CP}}(\mathit{c}\rightarrow \mathit{t})$,
which also determines the number of isomers in steady state: $R= N_c(t \rightarrow \infty) / N_t ( t \rightarrow \infty)$,
as discussed in Sec.\ \ref{sec: molecular model}. 

\begin{figure}
\begin{center}
\includegraphics{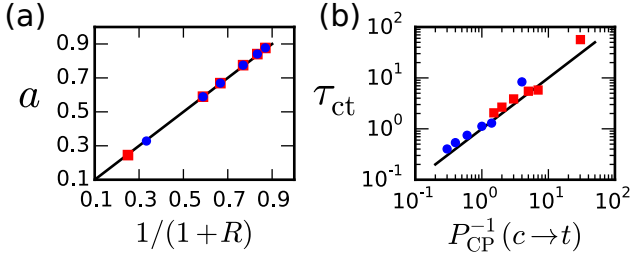}
\caption{(colors)
(a) The fit parameter $a$ in Eq.~(\ref{eq: trans number fit CP erasure}), which gives the relative number 
of \textit{trans} isomers 
in steady state, plotted versus $1/(1+R)$.
(b) The \textit{cis}-to-\textit{trans} 
relaxation
time $\tau_{ct}$ 
plotted
versus the inverse isomerization probability $P_{\mathrm{CP}}^{-1}(c \rightarrow t)$.
In both figures red squares are  
from simulations
with $P_{\mathrm{CP}}(t \rightarrow c)=10^{-1}$ and blue circles 
for $P_{\mathrm{CP}}(t \rightarrow c)=5.0 \cdot 10^{-1}$. The 
straight
black lines have zero $y$-intercept and unit slope.  
}
\label{fig: CP number trans fit}
\end{center}
\end{figure}

The implementation of our kinetic Monte-Carlo simulations suggests that the number of \textit{trans} isomers evolves according to the
following kinetic equation,
\begin{equation}
\frac{d \,  N_{t}(t)}{d \, t}=-\frac{1}{\tau_{tc}}N_t(t)+\frac{1}{\tau_{ct}}N_c(t) \, ,
\label{eq: trans number evolution}
\end{equation}
where $\tau_{ct}$ and $\tau_{tc}$ are 
characteristic 
relaxation times and $N_c(t) = N - N_t(t)$. It is solved by
\begin{equation}
N_t(t) / N =a+b \exp(- t / \tau)  \, ,
\label{eq: trans number fit CP erasure}
\end{equation}
with
\begin{equation}
a=\frac{1}{1 + \tau_{ct} / \tau_{tc}} \; , \; \, a+b=\frac{N_t(0)}{N} \; , \; \mathrm{and} \; \frac{1}{\tau} = \frac{1}{\tau_{tc}} + \frac{1}{ \tau_{ct}} \,
\end{equation}
Since 
$a$ is the steady-state value at $t \rightarrow \infty$, we also have $a=1/(1+R)$, which is confirmed in Fig.~\ref{fig: CP number trans fit}(a).
Thus, the ratio of isomerization probabilities also determines the ratio of the two relaxation times, 
$R= \tau_{ct}/\tau_{tc}$.
The fits of the simulated $N_t(t) / N$ in Figs.\ \ref{fig: CP erasure summary}(a) and (b) excellently confirm the kinetic model.
However, 
as in the case of thermal erasure,
we find 
$\tau_{ct} > P_{\mathrm{CP}}^{-1}(c \rightarrow t)$
[see Fig.\ \ref{fig: CP number trans fit}(b)], because in a crowded environment some of the attempted isomerization events are 
rejected.

In Fig.\ \ref{fig: CP erasure summary} we show the best fitting functions for $Q(t)$ as continuous lines. In both 
Figs.\ \ref{fig: CP erasure summary}(a) and (b) the relaxation of the birefringence is  exponential for $R<0.7$, stretched-exponential 
for $0.7 \le R < 0.3 $, and follows a
power-law for $R \le 0.3$. The characteristic times of the relaxation are shown as diamond markers and 
the exponents are given in Table~\ref{tab: exponent}.

As expected, a larger isomerization rate $P_{\mathrm{CP}}(t \rightarrow c)$ shifts the birefringence relaxation to smaller times
[compare Figs.\ \ref{fig: CP erasure summary} (a) and (b)] because
aligned \emph{trans} molecules are faster transformed to the \emph{cis}
state. Unlike the case of thermal erasure, the steady value of $N_t (t)/N$ is reached well before stretched-exponential or power-law relaxation
sets in. The relevant characteristic times $\tau_t$ are indicated by diamond markers
in Fig. \ref{fig: CP erasure summary}.
Interestingly, 
for constant $P_{\mathrm{CP}}(t \rightarrow c)$
the 
ratio of isomerization rates
$R$
controls
the functional form of the relaxation. 
In the power-law regime, the characteristic times 
$\tau_t$
do not change significantly,
while the
power-law exponent $\eta$
heavily depends on $R$
(see Table~\ref{tab: exponent}). 

At a first glance, the behavior in Fig.~\ref{fig: CP erasure summary} seems surprising. For increasing $R$ the isomerization rate
$P_{\mathrm{CP}}(\mathit{c}\rightarrow \mathit{t}) = P_{\mathrm{CP}}(\mathit{t}\rightarrow \mathit{c}) / R$ decreases and consequently 
the molecular orientations after isomerization become randomized less frequently. So, we expect the birefringence relaxation to become 
slower in contrast to the results presented in Fig.~\ref{fig: CP erasure summary}. However, we know already from thermal erasure 
that \emph{cis} isomers create a uniform environment, where the orientation of
\emph{trans} molecules relaxes faster and exponentially. This is the case for $R>1$, where the \textit{cis} isomers are in the majority
in steady state. Decreasing $R$ increases the number of rod-like \emph{trans} molecules, $N_t$. As discussed already for thermal erasure, 
they hinder the orientational relaxation of their neighbors more efficiently than \emph{cis} molecules. 
But they also create a more heterogeneous 
environment, where the birefringence relaxation first follows a stretched exponential and then for further decreasing $R$ becomes a 
power law. However, we were not able to quantify the dynamic heterogeneity using the dynamical susceptibility of Eq.~(\ref{eq: chi four}) 
for a subset of molecules as in Sec.\ \ref{subsec: thermal relaxation} 
since 
they continuously cycle between their two configurations.

\begin{table}[t]
\begin{center}
  \renewcommand{\arraystretch}{1.5}
  \begin{tabular}{ | c | c | c | c | c | c | c |}
    \cline{0-2}
    \cline{5-7}
    $P_{\mathrm{CP}}(\mathit{t}\rightarrow \mathit{c})$ & $R$ & $\beta$ & \; \; & $P_{\mathrm{CP}}(\mathit{t}\rightarrow \mathit{c})$ & $R$ & $\eta$ \\
    \cline{0-2}
    \cline{5-7}
    
    $5.0 \cdot 10^{-1}$ & 0.7 & 0.78 & &
    $5.0 \cdot 10^{-1}$ & 0.3 & 1.37 \\ \cline{0-2} \cline{5-7}
    
    $5.0 \cdot 10^{-1}$ & 0.5 & 0.77 & &
    $5.0 \cdot 10^{-1}$ & 0.2 & 0.95\\ \cline{0-2} \cline{5-7}
    
	$ 10^{-1}$ & 0.7 & 0.84 &  &
	$5.0 \cdot 10^{-1}$ & 0.15 & 0.82\\ \cline{0-2} \cline{5-7}
	
	$ 10^{-1}$ & 0.5 & 0.81 &  &
	$ 10^{-1}$ & 0.3 & 1.31\\ \cline{0-2} \cline{5-7}
	
	\multicolumn{3}{c}{} & &
	$ 10^{-1}$ & 0.2 & 1.10\\  \cline{5-7}
	
	\multicolumn{3}{c}{} & & 
	$ 10^{-1}$ & 0.15 & 0.71\\  \cline{5-7}
  \end{tabular}
  
\end{center}
\caption{Fit parameters for the 
%continuous 
solid
lines in Fig.\ \ref{fig: CP erasure summary}. Table on the left gives the exponent for the 
stretched-exponential fit (green lines) and table on the right gives the exponent for the power-law fit (red lines). }
\label{tab: exponent}
\end{table}

% % % % % % % % % % % % % % % % % % % % % % % % % % % % % % % % % % % % % % % % % % % % % % % % % % % % %
% % % thermal vs CP
% % % % % % % % % % % % % % % % % % % % % % % % % % % % % % % % % % % % % % % % % % % % % % % % % % % % %
\subsection{\label{sec: Thermal-CP comparision} Comparison between 
thermal
and CP erasure}

\begin{figure}[t]
\begin{center}
\includegraphics{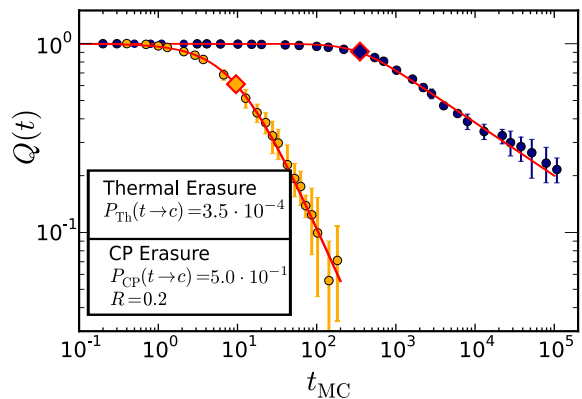}
\caption{(colors) Comparison of the birefringence relaxation during both CP and thermal erasure as obtained by 
kinetic Monte Carlo simulations. 
Solid
red lines 
are fits to
the power-law 
decay
of Eq.~(\ref{eq: PSSG relaxation}). The characteristic times of the 
power-law relaxation are shown as diamond markers. 
For both curves $\Temp=4$ and $\rho=20$. The
isomerization probabilities are given in the legend. Compare with Fig.\ \ref{fig: experiment scheme}.
}
\label{fig: thermal versus CP}
\end{center}
\end{figure}

In Fig.\ \ref{fig: thermal versus CP} we show the power-law relaxation of $Q(t)$ for both thermal and CP erasure using typical parameters.
Comparing it with 
Fig.\ \ref{fig: experiment scheme} 
gives an idea of the degree of agreement between our model and the experimental results.
Birefringence
relaxation is efficiently 
accelerated by the isomerization cycles induced by illumination with CP light. 
A similar speedup was demonstrated in
Refs.~[\onlinecite{teboul2011isomerization}] and [\onlinecite{teboul2009isomerization}],
where
isomerization of an azo-dye embedded in a molecular matrix significantly 
increased translational
diffusion of 
surrounding molecules.
By tuning the isomerization probabilities 
for
CP erasure, we achieved a difference between the characteristic
times $\tau_t$
in the power-law decay
of approximately two orders of magnitude, in good agreement with the experimental results.

The 
larger
isomerization rates 
during CP erasure and the presence of
\textit{cis} isomers 
in steady state give a faster power-law relaxation with a larger
exponent $\eta$, again in qualitative agreement with the experimental results.
Our model also accounts for the smaller steady-state  value of $Q(t)$ after illumination with CP light 
as
reported in Ref.~[\onlinecite{fang2013athermal}].
This
is due to the presence of the nearly isotropic \textit{cis} molecules that destabilize orientational order.                     

% % % % % % % % % % % % % % % % % % % % % % % % % % % % % % % % %
% % % % % % % % % % % % % % % % % % % % % % % % % % % % % % % % %
% %
% % Section : conclusion
% %
% % % % % % % % % % % % % % % % % % % % % % % % % % % % % % % % %
% % % % % % % % % % % % % % % % % % % % % % % % % % % % % % % % % 

\section{\label{sec: conclusion}
Summary and conclusion}

Experiments on SAMs with tethered light-switchable dye molecules show power-law relaxation of initial 
birefringence during both thermal erasure and 
the faster 
CP erasure.
Despite its simplicity, the molecular model discussed in this paper is able to reproduce
the experimental results and to 
identify dynamic heterogeneity as the main cause for the power-law decay.

First, we studied a system of pure \emph{trans} molecules. Here,
the non-exponential, glass-like relaxation of the orien\-ta\-tio\-nal order inscribed in the monolayer emerges naturally 
at high densities
and low 
temperatures
due to the presence of dynamic heterogeneity. 
Rotational motion
develops a transient sub-diffusive regime upon cooling. At the same time, molecules with dynamics faster and slower than the average become spatially correlated. 
The spatial average 
over these regions with different 
orientational mobilities results in the power-law decay of birefringence.        

In a second step we included the possibility that the molecules can assume two different isomeric forms.
During thermal erasure, the nearly isotropic \textit{cis} isomers
create a uniform environment
because they do not align locally with their neighboring molecules. 
Hence,
the orientation of \emph{trans} molecules relaxes exponentially.
The
experimental power-law relaxation of  
birefringence is 
only
recovered if 
a sufficient
number of \textit{trans} isomers is 
present. They slow down orientational relaxation but also initiate the formation of dynamic 
heterogeneities as in the pure \emph{trans} system.

During CP erasure
light adsorption induces a fast isomerization cycle between \emph{cis} and \textit{trans} isomers and thereby the overall orientational relaxation
becomes faster in agreement with
experimental results. 
The functional form of the birefringence relaxation 
is
controlled by the ratio 
of the two
isomerization probabilities, 
which determine the number of \emph{trans} and \emph{cis} molecules in steady state.
As in thermal erasure, a larger number of \textit{cis} isomers speeds up the exponential birefringence relaxation,
whereas \emph{trans} isomers in the majority hinder relaxation and ultimately give rise to a
power-law decay. Finally, the presence and nearly isotropic 
%\pc{\st{shape of}} 
\textit{cis} isomers also explains the smaller steady-state value, which birefringence reaches during CP erasure.
All
these findings strongly suggest the possibility to change the monolayer dynamics by tuning the absorption properties of the molecules 
and
their geometrical shape.

To reproduce the power-law decay of birefringence in our model, we have to fine-tune the parameters 
%\pc{\st{although}} 
within a range of 
%\pc{\st{reasonable}} 
values that are experimentally reasonable. 
It remains to be demonstrated if this is due to our simplified model or a general feature of light-switchable
molecules. Future work should address this question by exploring the effect of more complex molecular geometries with more
realistic molecular interactions. Another promising direction for exploring how light can be used to control material properties
are light-switchable sufactants \cite{karthaus1996reversible,shin1999using,ichimura2000light,eastoe2005self}.
They accumulate at fluid interfaces. By switching locally between the two isomeric states, 
the surface tension changes and its gradient drives Marangoni flow. This moves emulsion droplets along a surface or in bulk
with interesting non-linear dynamics \cite{diguet2009photomanipulation,Schmitt16}.


\begin{acknowledgments}

This work was supported by the Deutsche Forschungsgemeinschaft through the international research training group IRTG 1524. PC is thankful for support from the National Science Foundation Research Triangle Materials Research Science and Engineering Center (DMR-1121107).
\end{acknowledgments}


\bibliography{Bibliography}

\end{document}